\documentclass[aps, prl, twocolumn, letterpaper, nofootinbib, superscriptaddress, balancelastpage, longbibliography]{revtex4-1}
\usepackage[T1]{fontenc}
\usepackage[utf8]{inputenc}
\usepackage{graphicx}
\graphicspath{{figs/}}
\usepackage{caption}
\usepackage{subcaption}
\usepackage{amsmath, amssymb}
\usepackage[colorlinks,citecolor=blue,urlcolor=blue,linkcolor=blue]{hyperref}
\usepackage[dvipsnames]{xcolor}
\usepackage{makecell}

\begin{document}

\title{}

\author{Mao Zeng}
\email{mao.zeng@ed.ac.uk} 
\affiliation{Higgs Centre for Theoretical Physics, School of Physics and
Astronomy, University of Edinburgh, EH9 3FD, UK}

\date{\today}
\title{Reinforcement Learning and Metaheuristics for Feynman Integral Reduction}
\begin{abstract}
  We propose new methods for optimizing the integration-by-parts (IBP)
  reduction of Feynman integrals, an important computational
  bottleneck in modern perturbative calculations in quantum field
  theory. Using the simple example of one-loop massive bubble
  integrals, we pose the problem of minimizing the number of
  arithmetic operations in reducing a target integral to master
  integrals via the Laporta algorithm. This is a nontrivial
  combinatorial optimization problem over the ordering of IBP equation
  generation (from pairs of seed integrals and IBP operators) and the
  ordering of integral elimination. Our first proposed method is
  reinforcement learning, which involves an agent interacting with an
  environment in a step-by-step manner and learning the best actions
  to take given an observation of the environment (in this case, the
  current state of the IBP reduction process). The second method is
  using metaheuristics, e.g.\ simulated annealing, to minimize the
  computational cost as a black-box function of numerical priority
  values that control the orderings. For large-scale problems, the
  number of free parameters can be compressed by using a small neural
  network to assign priority values. Remarkably, with almost no human
  guidance, both methods lead to IBP reduction schemes that are
  competitive with the most efficient human-designed algorithms. We
  also found interpretable features in the AI results that may be
  applicable to more complicated problems.
\end{abstract}

\maketitle

\textbf{\textit{Introduction---}} Perturbative quantum field theory as
a precision science requires the evaluation of Feynman integrals from
multi-loop amplitudes and correlation functions, with many
applications in e.g.\ collider physics, statistical mechanics,
gravitational wave physics and cosmology. Integration-by-parts (IBP)
reduction \cite{Chetyrkin:1981qh} is used in almost all nontrivial
modern calculations except for those in highly symmetric special
theories, and is frequently the most computationally demanding part of
the calculations. IBP reduction is commonly carried out using the
Laporta algorithm \cite{Laporta:2000dsw} as implemented in publicly
available computer programs \cite{Anastasiou:2004vj, Studerus:2009ye,
  Lee:2013mka, Smirnov:2008iw, Smirnov:2013dia, Smirnov:2014hma,
  Smirnov:2019qkx, Smirnov:2023yhb, Maierhofer:2017gsa,
  Maierhofer:2018gpa, Klappert:2020nbg, Lange:2025fba}. While there
are many proposed alternative methods for IBP reduction, e.g.\ Refs.\
\cite{Lee:2012cn, Tarasov:2004ks, Gerdt:2005qf, Smirnov:2005ky,
  Smirnov:2006tz, Smirnov:2006wh, Lee:2008tj, Barakat:2022qlc,
  Gluza:2010ws, Schabinger:2011dz, Ita:2015tya, Larsen:2015ped,
  Abreu:2017xsl, Abreu:2017hqn, Bohm:2018bdy, Bendle:2019csk,
  Wu:2023upw, Wu:2025aeg, Mastrolia:2018uzb, Frellesvig:2020qot,
  Liu:2018dmc, Guan:2019bcx, Kosower:2018obg, Feng:2024qsa}, the
Laporta algorithm remains arguably the most widely-applied in
practical calculations. While the Laporta algorithm traditionally
required complicated manipulations of polynomials and rational
functions in intermediate steps, the use of finite-field numerical
techniques \cite{Kant:2013vta, vonManteuffel:2014ixa, Peraro:2016wsq,
  Abreu:2018zmy, Klappert:2019emp, Peraro:2019svx, Laurentis:2019bjh,
  Klappert:2020aqs, DeLaurentis:2022otd, Magerya:2022hvj,
  Belitsky:2023qho, Chawdhry:2023yyx, Liu:2023cgs, Maier:2024djk}
allows reconstructing complicated analytic results using a large
number of numerical evaluations. With such a workflow, it is important
to make each numerical IBP reduction run as fast as possible.

Applications of the Laporta algorithm involve many heuristic choices
that are based on experience rather than derived from first
principles, and adjusting such choices can have a tremendous impact on
the performance of IBP reduction. For example, there is a choice about
whether Lorentz-invariance relations \cite{Gehrmann:1999as} are
included, and the selection of seed integrals and IBP operators can be
optionally trimmed using Lie algebra relations
\cite{Lee:2008tj}. Recently, several papers observed that widely used
algorithms for selecting seed integrals can be improved to vastly
reduce the number of IBP equations \cite{Driesse:2024xad,
  Guan:2024byi, Bern:2024adl}.

The goal of this \textit{letter} is to present automated strategies
for \emph{blind} searches, with minimal human input, of IBP reduction
schemes that outperform, and/or provide valuable feedback on,
human-designed algorithms. Among recent interests in machine learning
in perturbative scattering amplitudes \cite{Dersy:2022bym,
  Jinno:2022sbr, Calisto:2023vmm, Cai:2024znx, Cheung:2024svk}, two
papers \cite{vonHippel:2025okr, Song:2025pwy} applied genetic
algorithms and LLM-powered program searches to find optimal symbolic
expressions controlling the various ordering choices in the Laporta
algorithm. Our work takes several novel directions not explored in the
previous literature. First, we initiate the study of reinforcement
learning in the context of Feynman integrals, which breaks down the
IBP reduction problem to a step-by-step decision process. Second, in
parallel, we introduce metaheuristics for minimizing the cost function
over numerical, rather than symbolic, priority values that control the
orderings. This provides a complementary perspective and allows
leveraging the vast body of available numerical optimization
algorithms. Third, we go beyond the binary inclusion
\cite{vonHippel:2025okr} and ordering \cite{Song:2025pwy} of seed
integrals, and take a fine-grained approach of ordering
\emph{seed-operator pairs}, which allows more aggressive
optimizations, e.g.\ applying only a subset of IBP operators to a
given seed integral \cite{Lee:2008tj, Smirnov:2013dia}. Finally, we
use the \emph{number of arithmetic operations}, tracked by our custom
IBP reduction program, as a more accurate proxy of computational
complexity than the number of seed integrals considered in previous
papers.

\begin{figure}
  \centering
  \includegraphics[width=0.2\textwidth]{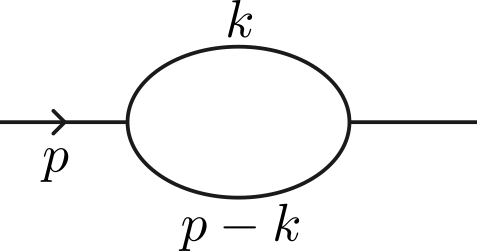}
  \caption{One-loop massive bubble integrals.}
  \label{fig:bubble}
\end{figure}

\textbf{\textit{IBP reduction \& simplified cost model---}} Consider the
family of one-loop massive \emph{bubble integrals} used as an introductory
example in the book \cite{Weinzierl:2022eaz}:
\begin{equation}
  \label{eq:bubble}
  I_{\nu_1, \nu_2} = \int \frac {d^D k}{i \pi^{D/2}} \frac{1}
  {(-k^2+m^2)^{\nu_1} [-(p-k)^2+m^2]^{\nu_2}} \, ,
\end{equation}
where the two propagators in Fig.~\ref{fig:bubble} are raised to
non-negative integer powers $\nu_1$ and $\nu_2$. The integral vanishes
in dimensional regularization if both indices are non-positive. When
one index is non-positive, the integral is referred to as a
\emph{tadpole integral}, otherwise a genuine bubble integral.
Integration-by-parts (IBP) equations arise as total derivatives
integrate to zero in dimensional regularization:
\begin{equation}
  \label{eq:ibp}
  0 = \int \frac {d^D k}{i \pi^{D/2}} \frac {\partial}
  {\partial k^\mu} \frac{q^\mu} {(-k^2+m^2)^{\nu_1}
    [-(p-k)^2+m^2]^{\nu_2}} \, ,
\end{equation}
where $q^\mu$ is any Lorentz vector. With a slight abuse of notation,
we refer to the above IBP equation as the outcome of applying the
\emph{IBP operator}, $(\partial / \partial k^\mu) q^\mu$, to the
\emph{seed integral}, $I_{\nu_1, \nu_2}$. Specifically, we have
\begin{itemize}
\item IBP operator 1: $(\partial / \partial k^\mu) k^\mu$.
\item IBP operator 2: $(\partial / \partial k^\mu) p^\mu$.
\end{itemize}
We only study IBP relations and disregard symmetry relations
$I_{\nu_1, \nu_2} = I_{\nu_2, \nu_1}$ for simplicity. It turns out
that all $I_{\nu_1, \nu_2}$ can be reduced to three \emph{master
  integrals}, $I_{1,1}, \, I_{1,0}, \, I_{0,1}$.
Now we walk through an application of the Laporta algorithm to reduce
the \emph{target integral} $I_{1,2}$ to the three master integrals at
numerical values of dimension and kinematic variables,
\begin{equation}
  \label{eq:numericalKinematics}
  p^2=1, \quad m^2=2, \quad D=5/7 \, ,
\end{equation}
which allows us to illustrate a simple cost model based on the number
of arithmetic operations. The numerical values are unimportant, since
the number of arithmetic operations is usually the same at different
numerical values as long as the IBP system remain non-singular, with
occasional small differences due to accidental cancellations. An
example sequence of IBP reduction steps, alternating between (a)
generating an equation and reducing against all previous rules, and
(b) eliminating an integral to form a reduction rule, is:
\begin{itemize}
\item Step 1a: generate an IBP equation from seed integral $I_{1,1}$
  and IBP operator 2,
  \begin{equation}
    -I_{0,2} + I_{2,0} - I_{1,2} + I_{2,1} = 0 \, .
  \end{equation}
  We consider equation generation to be cheap, so do not track its
  cost. Now we reduce the above equation against previous reduction
  rules; currently there are none since we are at step 1, so there is
  no work to do and no \textbf{row reduction cost} (considering linear
  equations as rows in a matrix) is incurred.
\item Step 1b: Choose an integral to eliminate. We choose $I_{1,2}$ to
  produce the reduction rule,
  \begin{equation}
    \label{eq:rule1}
    I_{1,2} \rightarrow -I_{0,2} + I_{2,0} + I_{2,1} \, .
  \end{equation}
  In this step, we had to normalize the coefficient of the eliminated
  integral on the LHS of the reduction rule to $1$, and we count the
  \textbf{normalization cost} to be 4, i.e.\ the length of the
  equation. (In this case, the normalization is trivial, i.e.\
    a sign flip, but this is a coincidence from our simple choice of
    kinematic variables Eq.~\eqref{eq:numericalKinematics}.)
\item Step 2a: generate an IBP equation from seed integral $I_{1,1}$
  and IBP operator 1,
  \begin{equation}
    -I_{0,2} - (16/7) I_{1,1} + 3 I_{1,2} + 4 I_{2,1} = 0 \, .
  \end{equation}
  Reducing against all previous reduction rules, in this case the
  single rule Eq.~\eqref{eq:rule1}, yields the reduced IBP equation,
  \begin{equation}
    -4 I_{0,2} + 3 I_{2,0} - (16/7) I_{1,1} + 7 I_{2,1} = 0 \, .
  \end{equation}
  Since we reduced against a previous rule of length 4 (counting both
  LHS and RHS), we record a row reduction cost of 4. Generally, we
  need to reduce against more than one previous rule, and the cost is
  the sum of the lengths of all rules applied.
\item Step 2b: Choose an integral to eliminate. We choose $I_{2,1}$ to
  produce the reduction rule,
  \begin{equation}
    \label{eq:rule2}
    I_{2,1} \rightarrow (4/7) I_{0,2} - (3/7) I_{2,0} + (16/49) I_{1,1} \, ,
  \end{equation}
\end{itemize}
again with normalization cost $4$.
The chain of rules, Eq.~\eqref{eq:rule1} and \eqref{eq:rule2}, reduces
the target integral $I_{1,2}$ to the master integral $I_{1,1}$ and
tadpole integrals. The cost incurred so far is 12, and continuing the
process with tadpole seed integrals will finish the job of fully
reducing $I_{1,2}$ the three master integrals.

\textbf{\textit{Reinforcement learning---}} We start with some brief
definitions. Reinforcement learning (RL) \cite{barto2021reinforcement}
is a branch of machine learning formulated in terms of Markov decision
processes (MDPs), where an \emph{agent} interacts with an environment
and takes \emph{actions} based on \emph{observations} of the
environment state. The actions cause transitions to other states and
possibly positive feedback signals called \emph{rewards}. When
\emph{terminal states} are reached, e.g.\ upon completion of some
goal, a complete \emph{episode} of interactions comes to an end.  In
\emph{deep reinforcement learning}, specifically variants based on
``policy gradients'', a deep neural network is trained to learn an
optimized \emph{policy}, i.e.\ a map between observations to the
probability of taking each action, to maximize the expected future
rewards. See Ref.~\cite{Constantin:2021for} for a previous application
of RL to theoretical high-energy physics.

To apply RL, we first specify the rewards. The reward received after
each step is the negative of the cost (of row reduction or
normalization) as illustrated in the previous section, until the
episode terminates when we have succeeded in reducing $I_{3,3}$ to the
three master integrals. Therefore, all the rewards in our environment
are non-positive, and maximizing the reward is equivalent to
minimizing the cost. In principle, we could assign a positive reward
for achieving the goal of reducing the target integral to master
integrals, but since we have a finite system which always terminates
with achieving this goal, such a positive reward would become an
uninteresting constant for every episode.

\begin{figure}[t]
    \centering
    \begin{subfigure}[b]{0.24\textwidth}
        \includegraphics[width=\textwidth]{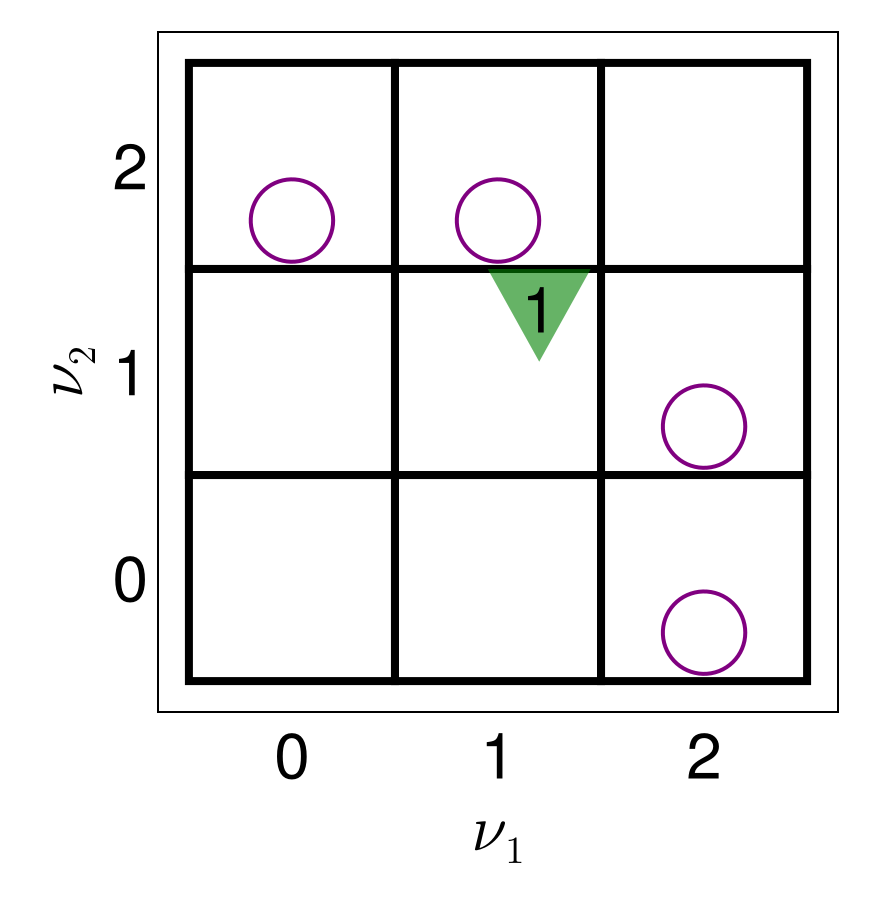}
        \caption{}
    \end{subfigure}
    \begin{subfigure}[b]{0.24\textwidth}
        \includegraphics[width=\textwidth]{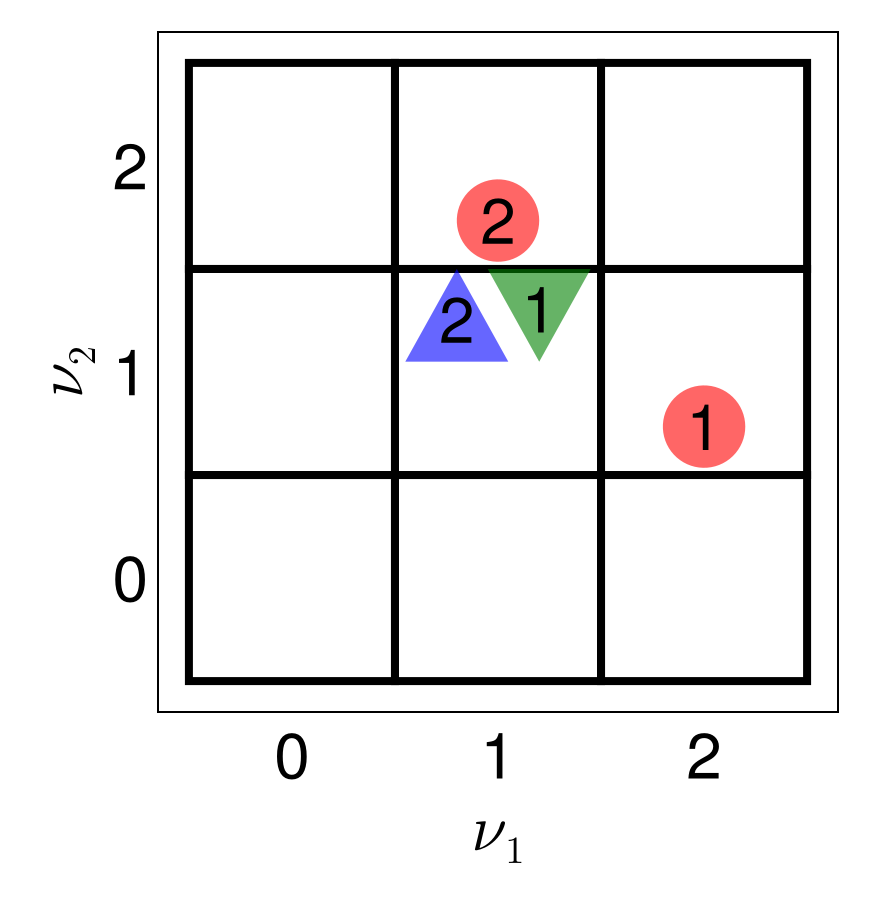}
        \caption{}
    \end{subfigure}
    
    
    \caption{Graphical representations of the state of the IBP
      reduction process for the target integral $I_{1,2}$ after steps
      (1a) and (2b).}
    \label{fig:observations}
  \end{figure}

\begin{figure}
  \centering
  \includegraphics[width=0.36\textwidth]{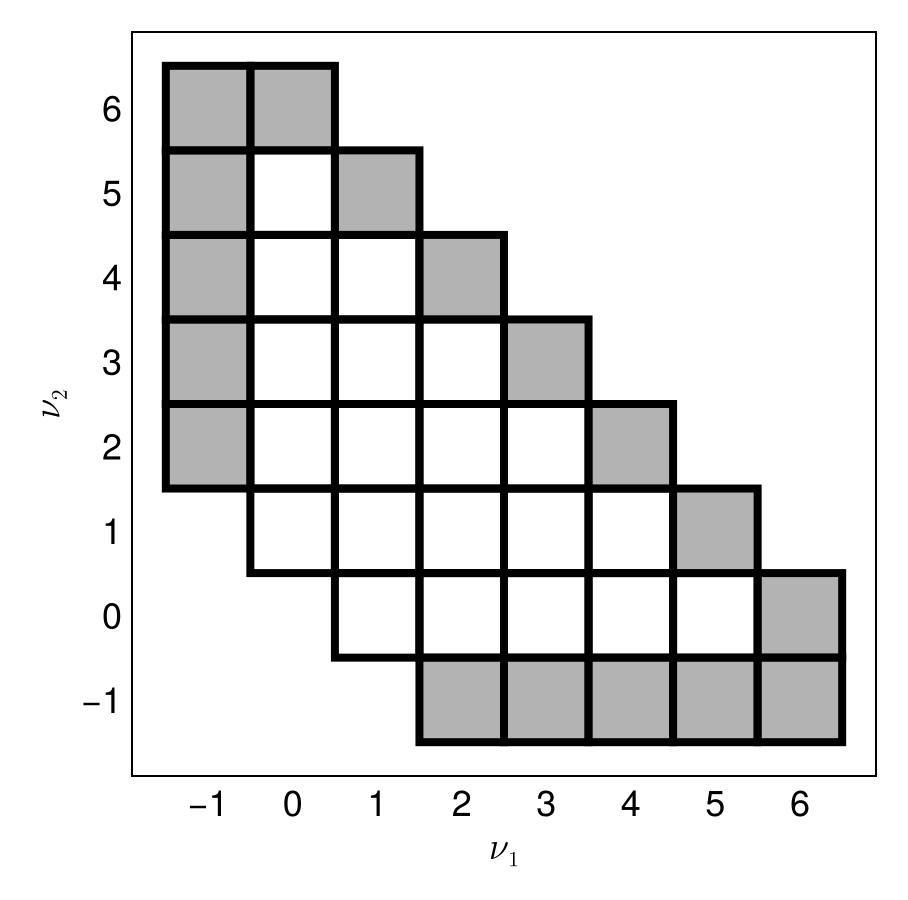}
  \caption{The $20$ candidate seed integrals (white squares) and $17$
    additional auxiliary integrals (light gray squares on the
    peripherals).}
  \label{fig:emptyboard}
\end{figure}

Then we specify the observations. let us first revisit the IBP
reduction of $I_{1,2}$ in the previous section before moving on to
more nontrivial examples.  We create graphical representations of the
steps 1a and 2b, as two examples, in Fig.~\ref{fig:observations}, which
extends Fig.~\ref{fig:emptyboard} with information about the IBP
reduction process. An upper triangle $\triangle$ (or a lower triangle
$\triangledown$) in a grid denotes that the seed integral has been
used, in the most recent step or any previous step, with the IBP
operator 1 (or IBP operator 2). The number in the triangle denotes the
step at which a seed-operator is used to generate an IBP
equation. Unfilled circles denote the integrals in the last reduced
IBP equation (excluding the three master integrals) that are
candidates to be eliminated to produce a new reduction rule. A filled
circle denotes that an integral has been eliminated, and the number in
it denotes the step at which this happened.
  
To target a more nontrivial integral $I_{3,3}$, we enlarge the board
to have 37 squares in Fig.~\ref{fig:emptyboard}, where white square
are candidate seed integrals, and gray squares are extra integrals
that may appear in IBP equations.
The presence/absence and the step number in the three graphical
features (upper/lower triangle, circle) are encoded in a
$37\times 3 =111$ dimensional numerical vector as observation data.
We use the simplest neural network architecture, a multi-level
preceptron (MLP), which takes the $111$ numbers as input and process
the data through 3 hidden layers, each with $256$ neurons, before
generating an output whose dimension is also $111$, for probabilities
for the next action. (The actual input dimension of nonzero data is
$77$, since the gray squares in Fig.~\ref{fig:emptyboard} are not used
as seed integrals. Similarly, the relevant output dimension for
actions is at most $77$, since the action must alternate between
equation generation and elimination, and any seed-operator pair can
only be used once.)  We use the RL algorithm of proximal policy
optimization \cite{schulman2017proximal}. Implementation details are
covered by the supplemental material.

\begin{figure}[h]
  \centering
  \includegraphics[width=0.45\textwidth]{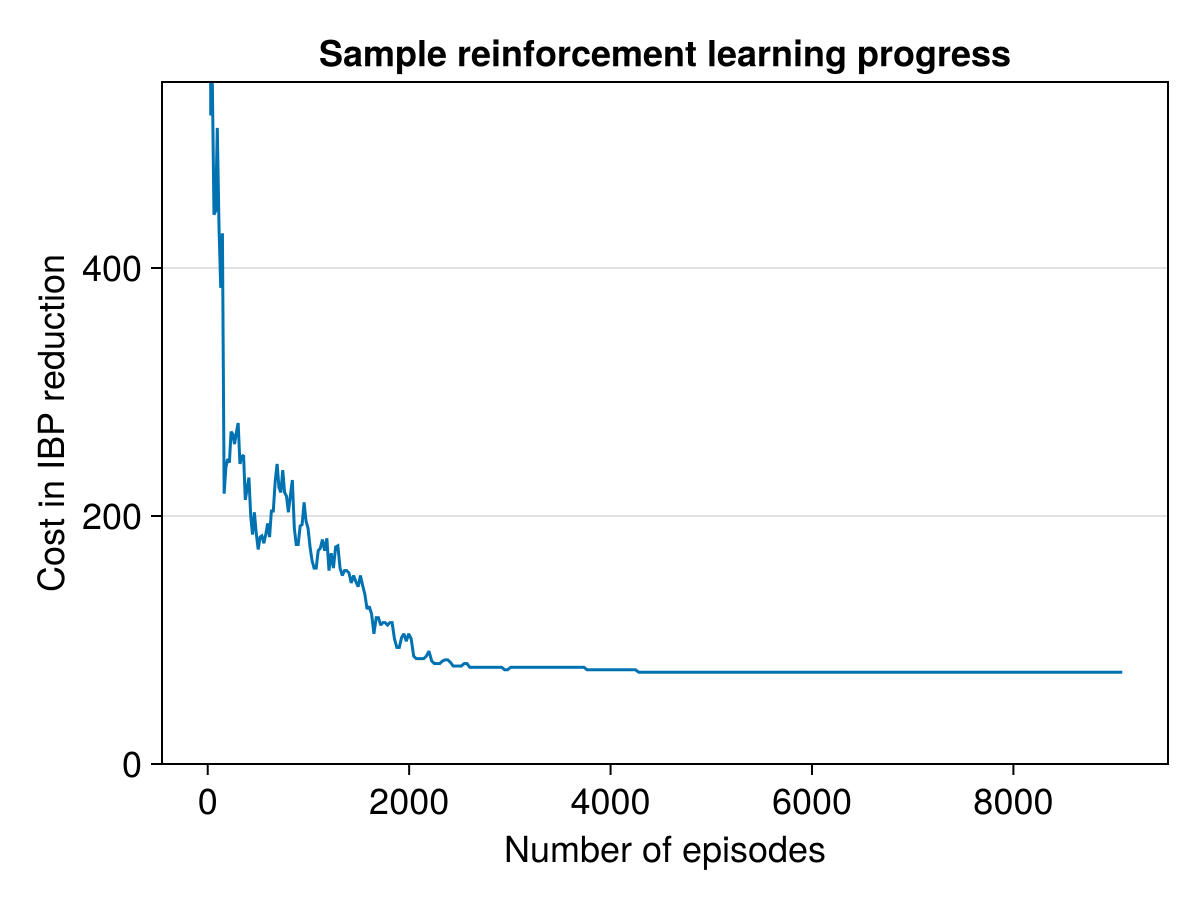}
  \caption{Cost in reducing $I_{3,3}$ to master integrals versus the
    number of episodes used for training, in an example run that
    reached the best cost 74 after 4280 episodes. The cost at the
    beginning of training is 1234, outside the plot range, but
    drops below 550 after 30 episodes.}
  \label{fig:rlcurve}
\end{figure}

\begin{figure}[h]
  \centering
  \includegraphics[width=0.45\textwidth]{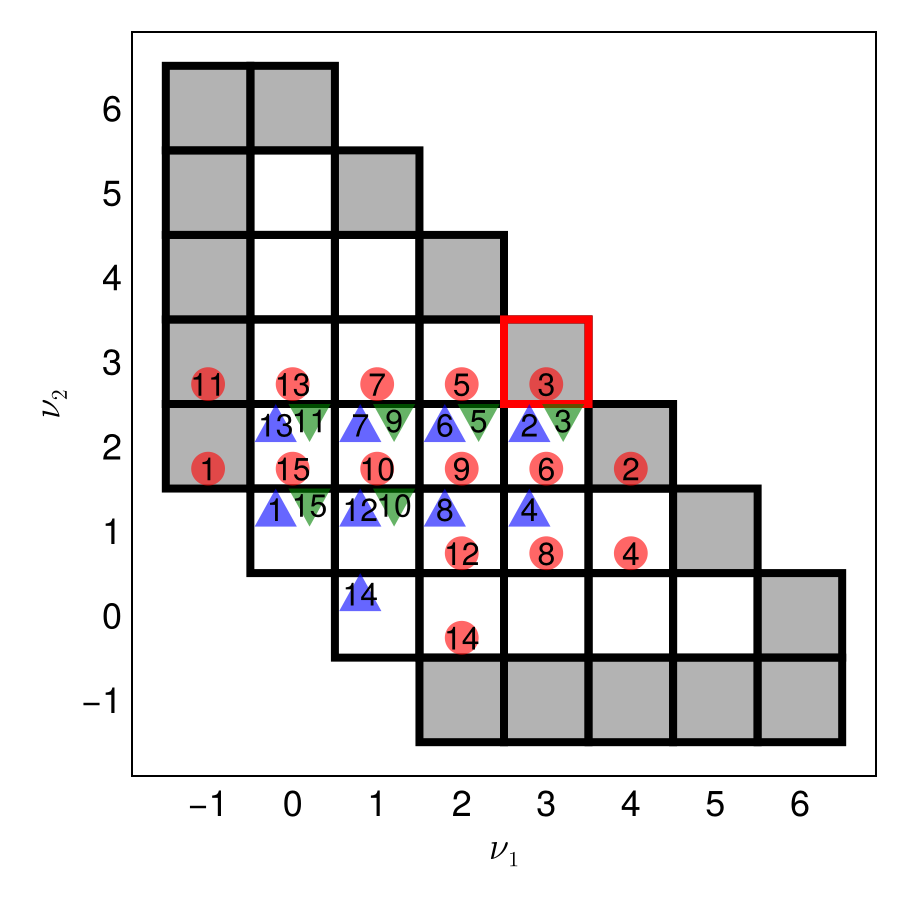}
  \caption{The optimal IBP reduction steps found by reinforcement
    learning, for reducing $I_{3,3}$ marked as the red square.}
  \label{fig:i33best}
\end{figure}

We run the algorithm 32 times, each with a maximum of 322560
steps. Typically a few dozen steps make up one episode. The costs
achieved at the end of the 32 training runs have a median of 93.5 a
minimum of 74.
The minimum cost is achieved by 2 of the runs, and the learning curve
showing the cost versus the number of episodes is in
Fig.~\ref{fig:rlcurve}.
The IBP reduction steps leading to the best cost of $74$ is shown in
our graphical notation in Fig.~\ref{fig:i33best}. Strikingly, starting
from random exploration, RL has produced an IBP
reduction scheme with distinct and interpretable features. First, the
choice of seed integrals (denoted by triangles) adheres to a
rectangular region with $\nu_1 \leq 3, \, \nu_2 \leq
2$. Second, the seed integrals are used in a strictly descending order
in $\nu_1+\nu_2$, ignoring the tadpole seed integrals. Third, elimination (denoted by
numbers in circles) also follows a strictly descending order in
$\nu_1+\nu_2$ when ignoring tadpole integrals. The machine-generated
log documenting the detailed steps is included in the supplemental
material.

\textbf{\textit{Metaheuristics---}} We propose a second method for
optimizing IBP reduction using the same cost model and action
space. We assign numerical priority values to the $40$ seed-operator
pairs corresponding to the white squares in Fig.~\ref{fig:emptyboard}
with either of the two IBP operators, which induce an ordering (in
descending priority values) for generating IBP equations. Similarly,
we assign numerical priority values to all $34$ non-master integrals
to order elimination. Combined, the $74$ priority values completely
specify the IBP reduction process, again for the example target
integral $I_{3,3}$. IBP reduction proceeds with the prescribed order
until a chain of reduction rules has been found which reduces the
target integral to master integrals, and at this point the process
terminates. The integer-valued IBP reduction cost is therefore a
function of the 74 priority values, and we can feed this
\emph{black-box function} to numerical minimization algorithms.

\begin{figure}[h]
  \centering
  \includegraphics[width=0.45\textwidth]{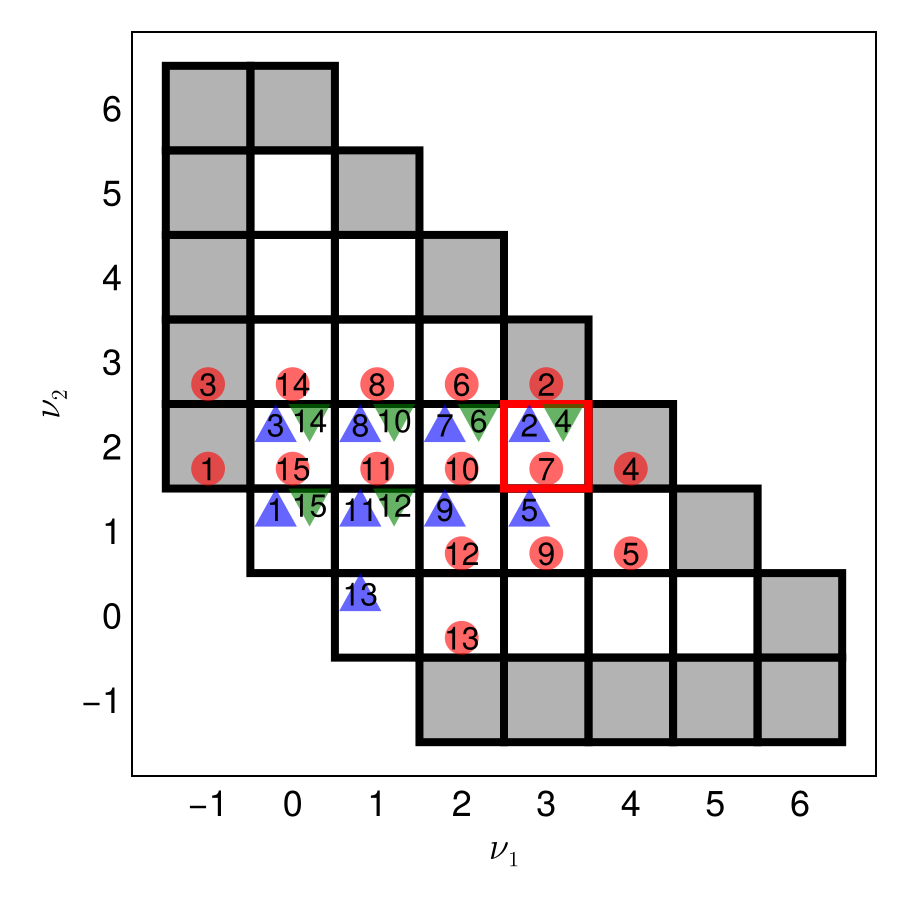}
  \caption{The optimal IBP reduction steps found by simulated
    annealing, for reducing $I_{3,3}$ marked as the red square.}
  \label{fig:i33best_sa}
\end{figure}

Given the non-differentiable (piecewise flat) nature of the cost
function, we use metaheuristics, i.e.\ high-level search strategies
for approximate optimal solutions. Specifically, we use
\emph{simulated annealing} (SA) \cite{aarts1987simulated}, and treat
the IBP reduction costs as energy levels in a thermodynamic system. We
start with a population of $200$ randomly initialized points in the
$74$-dimensional space of priority values bounded in the interval
$[-2,2]$. In each iteration, for each point, a random neighbor point
at a small distance is probed. A transition to the neighbor point
takes place if the latter has a lower cost, and if the cost is higher,
the transition can still take place but with a probability that is
suppressed by a Boltzmann factor dependent on the increase in
cost. The temperature is initially set high to encourage exploration,
and then gradually lowered to allow the points to settle down at
values that minimize the black-box function. We ran 32 times with
different random numbers, with about 50 iterations and 9602 function
evaluations in each run (comparable with the RL run in
Fig.~\ref{fig:rlcurve}). The lowest cost is again 74, same as obtained
with RL, found by 3 of the 32 SA runs. The median cost achieved is
$84.5$. The IBP reduction steps found by a run achieving the lowest
cost is illustrated in Fig.~\ref{fig:i33best_sa}.
We can see that the seed integrals used and the IBP operators used
with each seed integral are in complete agreement with the best RL
solution in Fig.~\ref{fig:i33best}, despite some differences in the
order of the steps.\footnote{Note that when two computation steps are
  independent of each other, swapping them trivially has no
  effect. Therefore, a careful dependency analysis, not carried out in
  this work, would be needed to actually establish whether the
  solutions of Figs.~\ref{fig:i33best} and \ref{fig:i33best_sa} are
  equivalent or not.}

\begin{figure}[h]
  \centering
  \includegraphics[width=0.3\textwidth]{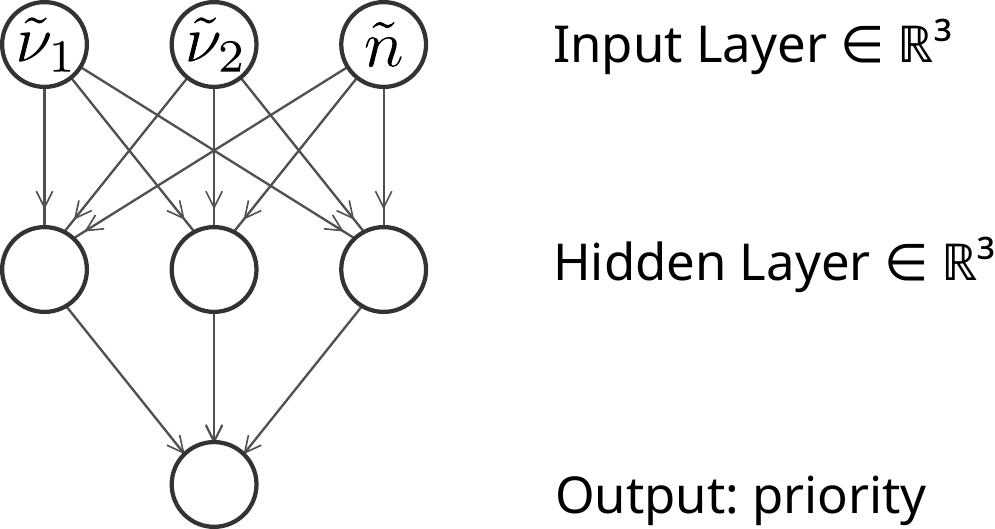}
  \caption{A tiny neural network for assigning priority values, with
    normalized inputs $\tilde \nu_i = \nu_i / \operatorname{max}(\nu_i)$,
    $\tilde n = n / \operatorname{max}(n)$.}
  \label{fig:nn}
\end{figure}

\textbf{\textit{Parameter compression with a neural network---}} Now
we propose a way to scale up the simulated annealing method to handle
large-scale problems without a prohibitively large increase in the
number of free parameters. We use a tiny neural network illustrated in
Fig.~\ref{fig:nn}, with 15 free parameters, to assign priority values
for the integral $I_{\nu_1, \nu_2}$, paired with an IBP operator
labeled $n>0$ for seeding or with $n=0$ for elimination. Constant
normalization factors are applied to rescale the three input numbers
to between 0 and 1.
Then we use simulated annealing again to optimize the neural network
parameters.

For $I_{3,3}$, we obtain a minimum cost of 121 and median cost of
162.5 in 32 runs. Though not as good as results from RL and SA with
the full 74 priority parameters, the minimum cost here still beats
common human-designed algorithms, as we will see later.

\begin{figure}[h]
  \centering
  \includegraphics[width=0.45\textwidth]{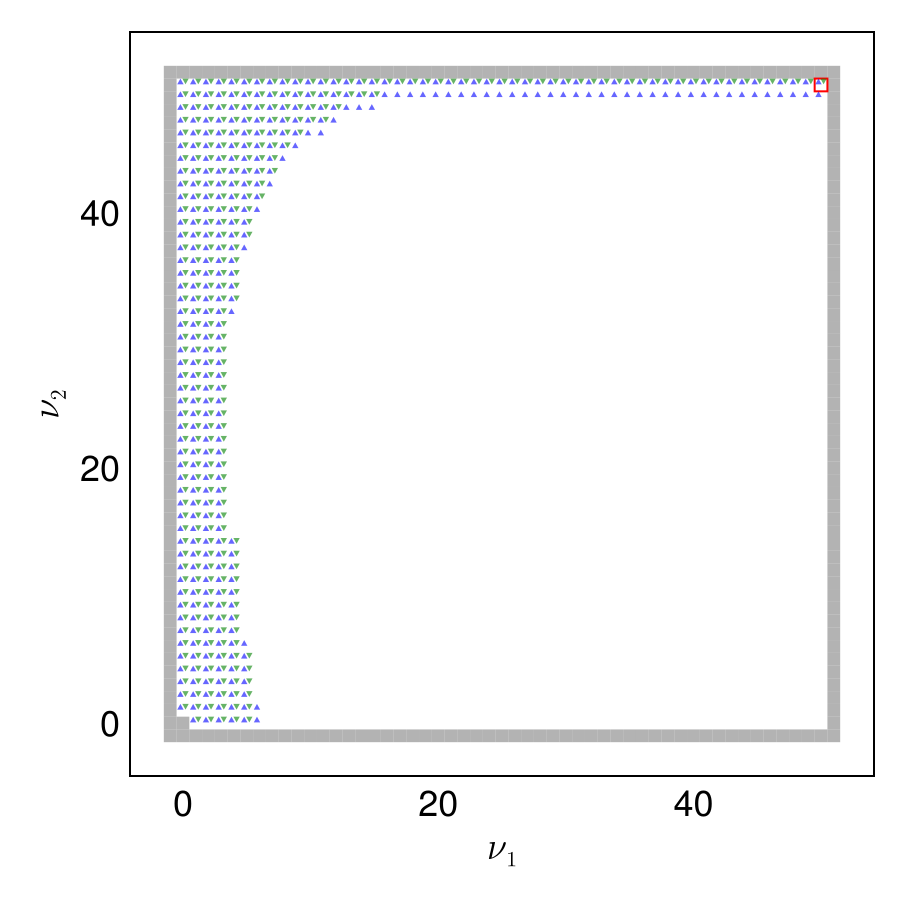}
  \caption{The optimal IBP reduction steps found by simulated
    annealing with parameter compression by a neural network, for
    reducing the target integral $I_{50,50}$. The tiny red square in
    the top right corner marks the target integral. The bottom left
    gray square corresponds to $(-1,-1)$. Auxiliary integrals not
    allowed to be seed integrals are colored gray. Grid lines are
    omitted to avoid cluttering the figure.}
  \label{fig:best50}
\end{figure}

We stress-test the method with the much harder problem of reducing
$I_{50, 50}$. In this case, we restrict the seed integral candidates
to a rectangular range $0 \leq \nu_{1,2} \leq 50$ rather than a larger
triangular range of the type in Fig.~\ref{fig:emptyboard} which
restricts $|\nu_1|+|\nu_2|$. Despite starting from a slightly more
informed choice that reduces the number of seed integral candidates by
one half, the automated algorithm still needs to discover the
nontrivial optimized orderings of seed-operator pairs and
elimination. The number of priority values is about 7500, which makes
it very difficult to apply SA directly, so the parameter compression
of the neural network becomes essential. The seed integral choice of
the best solution found in 32 runs is shown in Fig.~\ref{fig:best50},
A striking and interpretable feature of Fig.~\ref{fig:best50} is that
only a small part of the square region is used as seed integrals,
concentrated in two narrow strips, one at the top and one on the left,
showing a dramatic reduction of the number of seed integrals compared
with the human-designed rectangular seeding scheme
$-1 \leq \nu_1, \nu_2 \leq 50$, i.e.\ filling up the full area in the
figure. We refer to this as the ``double strip'' seeding scheme.

\begin{figure}[h]
  \centering
  \includegraphics[width=0.45\textwidth]{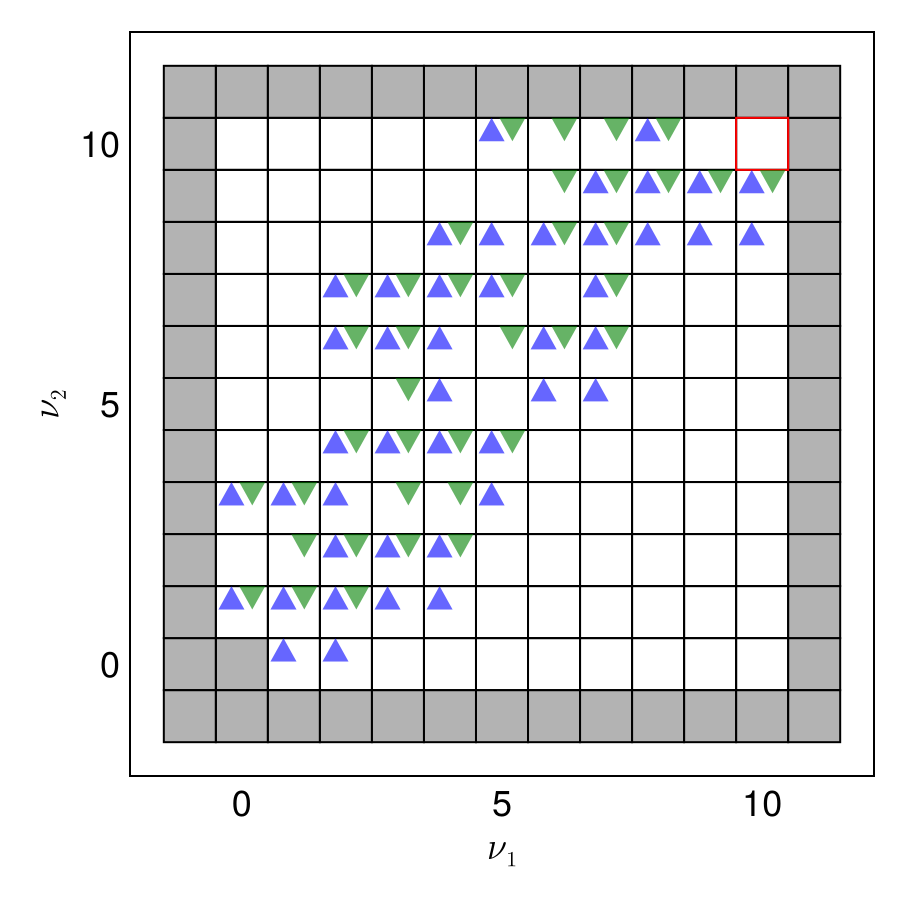}
  \caption{The optimal IBP reduction steps found by simulated
    annealing for the full priority values, i.e.\ without parameter
    compression, for reducing the target integral $I_{10,10}$ (marked
    as a red square).}
  \label{fig:best10}
\end{figure}

\begin{figure}[h]
  \centering
  \includegraphics[width=0.45\textwidth]{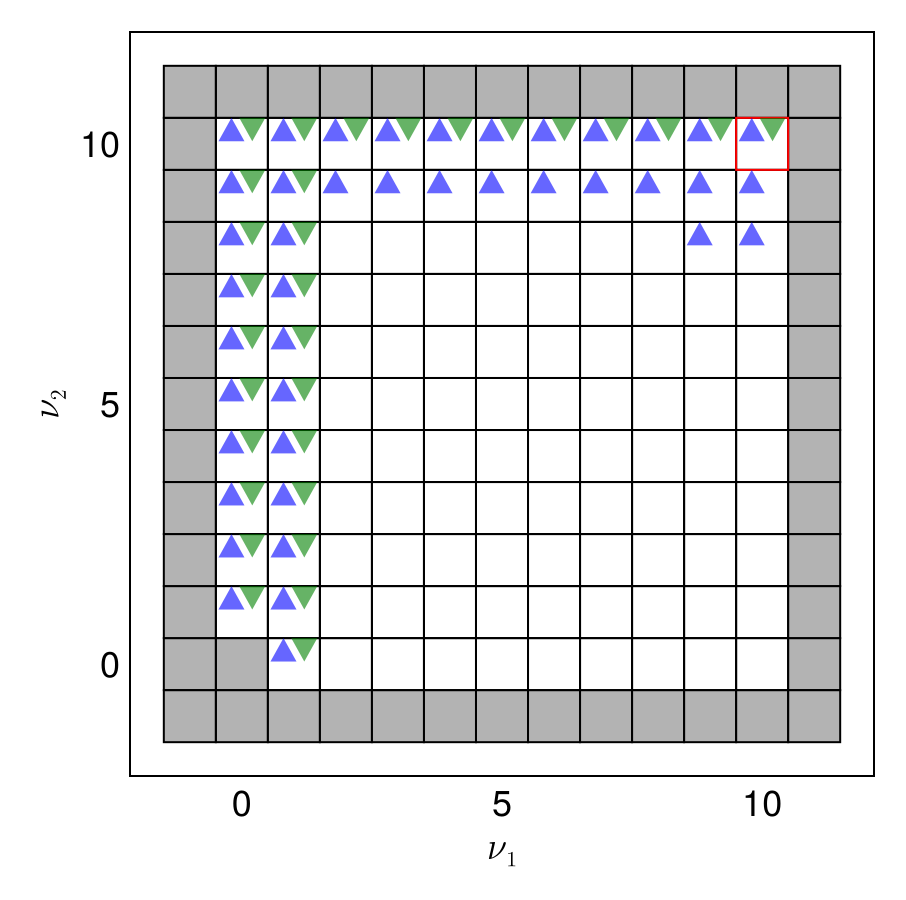}
  \caption{The optimal IBP reduction steps found by simulated
    annealing with parameter compression by a tiny neural network, for
    reducing the target integral $I_{10,10}$ (marked as a red
    square).}
  \label{fig:best10_nn}
\end{figure}

It is informative to look at another example, the reduction of
$I_{10,10}$, which is a moderately sized problem which is still small
enough to allow the application of simulated annealing both with and
without the parameter compression by a neural network, while
demonstrating more interesting features than the small-scale problem
of reducing $I_{3,3}$. In Fig.~\ref{fig:best10}, we show the
seed-operator choices found by a single SA run optimizing the full set
of priority values, and Fig.~\ref{fig:best10_nn} shows the choices
found by a single SA run with parameter compression. Each run starts
with 300 randomly initialized points and uses about 100,000 IBP
reduction runs. The two schemes found achieve costs of 961 (without
parameter compression) and 919 (with parameter compression),
respectively. Both figures omit the step numbers and the elimination
choices in the interest of simplicity.
Again, both schemes use significantly fewer seed integrals than the
human-designed rectangular seeding scheme and leave large blank areas
in the figures due to unused seed integrals. The scheme of
Fig.~\ref{fig:best10_nn} has a similar ``double strip'' structure as
Fig.~\ref{fig:best50} for the larger-scale $I_{50,50}$ problem but has
a more regular appearance, possibly because a larger number of steps
would need to be run for the $I_{50,50}$ problem to converge to the
same regular pattern.

\textbf{\textit{Comparison with human-designed algorithms---}} As a
baseline algorithm resembling those in widely available public IBP
codes, we follow Laporta \cite{Laporta:2000dsw} and generate IBP
equations first from tadpole seed integrals and then genuine bubble
seed integrals, chosen from the candidates indicated as white squares
in Fig.~\ref{fig:emptyboard}, ordered with ascending complexity of the
seed integrals. The complexity can be taken as $|\nu_1|+|\nu_2|$,
using $\nu_1$ to break ties. Each seed integral is used with the IBP
operator 1 followed by operator 2.  As a more optimized human-designed
algorithm, we further restrict seed integrals to the rectangular range
$\nu_1 \leq n_1, \nu_2 \leq n_2$ for reducing $I_{n_1,n_2}$. Optionally, we remove
redundant IBP equations, i.e.\ those linearly dependent on previous
equations, using the idea of Ref.~\cite{Kant:2013vta, Peraro:2019svx}
and discard the costs associated with the redundant equations.
Finally, as the most dedicated human optimization effort, for
$I_{3,3}$ only, we apply the advanced ordering algorithms of {\tt
  Kira} \cite{Maierhofer:2018gpa, Klappert:2020nbg, Lange:2025fba},
which requires generating all IBP equations at once for a global
analysis at the cost of higher initial memory usage, using separate
hand-tuned cutoffs for the two propagator powers following the recent
work of Ref.~\cite{Driesse:2024xad}.

Table \ref{tab:comparison} summarizes the IBP reduction costs achieved
for reducing $I_{3,3}$, $I_{10,10}$, and $I_{50,50}$ found from
reinforcement learning, two variants of simulated annealing, with or
without parameter compression using a tiny neural network, and
human-designed algorithms. Note that applying RL to problems larger
than $I_{3,3}$ is left to future work due to technical
limitations. For $I_{3,3}$, RL and SA both outperform most
human-designed algorithms and achieve parity with the most optimized
one with the aforementioned tradeoffs in memory usage and manual
tuning effort.

For $I_{50,50}$ and $I_{10,10}$, we have already seen, in
Figs.~\ref{fig:best50}, \ref{fig:best10}, \ref{fig:best10_nn}, that
significantly fewer seed integrals are used than human-designed
algorithms, and indeed incur lower reduction costs according to Table
\ref{tab:comparison}.
In the case of $I_{50,50}$ only, the steps found by SA contain
redundant equations, and simply truncating redundant equations further
improves the result.
\begin{table*}[t]
\centering
\begin{tabular}{|c|c|c|l|}
\hline
  Integral & $I_{3,3}$ & $I_{10, 10}$ & $I_{50, 50}$ \\ \hline
  RL & 74 & {} & {} \\ \hline
  SA & 74 & 919 &  {} \\ \hline
  SA + NN & 121 & 961 & $17843$ \\ \hline
  SA + NN + truncation & 121 & 961 & $12775$ \\ \hline
  Human (baseline) & 365 & 3951 & $92471$ \\ \hline
  Human (rectangular seeding + truncation) & 237 & 2239 & 43579 \\ \hline
  \makecell{Human (advanced ordering heuristics \\ + separate cutoffs
  for $\nu_1$ and $\nu_2$)} & 74 & {} & {} \\ \hline
\end{tabular}
\caption{Comparison of the best IBP reduction costs found by
  reinforcement learning (RL), simulated annealing (SA), possibly with
  a reduced parameter space using a neural network (NN), and
  human-designed algorithms. RL and SA runs depend on random number
  generators and are repeated for 32 times for $I_{3,3}$, with the
  lowest costs tabulated. Unlike the case of $I_{3,3}$, all
  $I_{10,10}$ and $I_{50,50}$ runs, including the baseline human
  algorithm, start from a rectangular seed range, possibly subject to
  further optimization by RL or SA algorithms.}
\label{tab:comparison}
\end{table*}

\textbf{\textit{Conclusion---}} IBP reduction of Feynman integrals is
an important computational bottleneck in difficult calculations in
perturbative QFT. We have demonstrated two new methods, namely
reinforcement learning and metaheuristics, for optimizing IBP
reduction by adjusting various orderings in the Laporta
algorithm. One-loop massive bubble integrals are used to illustrate
the methods. Despite the apparent simplicity of such integrals, the
problem is surprisingly nontrivial, and reinforcement learning and
metaheuristics both succeed in finding computational steps that are
competitive with the most efficient human-designed algorithms. In
fact, for the simplest example of reducing the bubble integral with
each propagator raised to a cubic power, the two methods agree with
each other in the lowest cost reached, and an example scheme that
achieves this lowest cost is documented in detail in the supplemental
material as a benchmark for future investigations. We also took
initial steps in tackling large-scale problems, again out-performing
common human-designed algorithms for reducing $I_{50,50}$, which
required tens of thousands of arithmetic operations. This was done by
using a tiny neural network to reduce the number of free parameters
and then applying simulated annealing. The application of
reinforcement learning to large-scale problems is left as an open
question for future studies. Since a first preprint version of the
paper appeared, a moderate-sized example of applying simulated
annealing to the reduction of $I_{10,10}$ has been added. The results
in Figs.~\ref{fig:best10_nn} and \ref{fig:best50} suggest a novel
``double-strip'' seeding scheme with apparent linear rather than
quadratic scaling of the number of seed integrals. It would be
interesting to see if such a scheme can be generalized to more
complicated families of integrals. Meanwhile, the result in
Fig.~\ref{fig:best10} has some resemblance to the ellipse scheme found
in Ref.~\cite{Song:2025pwy}, where seed integrals are concentrated in
a diagonal area. The interpretability of the AI results may also
inform better designs of algorithms by humans.

The two methods have their advantages and disadvantages. The ease of
implementation is a major advantage of using metaheuristics to
optimize the IBP reduction cost as a black-box function of the
orderings (encoded in numerical priority values). Though we focused on
simulated annealing inspired by thermodynamics, a vast body of other
metaheuristic algorithms, including particle swarm optimization and
evolutionary/genetic algorithms, can be applied in an essentially
plug-and-play manner. Reinforcement learning introduces the
perspective of dynamically adjusting the chosen actions by reacting to
changes in the environment. This brings the possibility that a single
agent can be trained to adjust and adapt to a large number of
different IBP reduction problems, e.g.\ involving different families
of Feynman integrals.

Since the methods presented, in the current state, require thousands
of IBP runs with different ordering choices before converging to
optimal ones, the most likely practical application would be to the
IBP reduction of Feynman integrals with a large number of kinematic
scales, which could require hundreds of thousands or even more
numerical finite-field runs for analytic results to be
reconstructed. The cost savings in such a large number of later runs
would make the initial optimization cost worthwhile. It is of course
highly desirable to improve our methods and achieve better
\emph{sample efficiency}, i.e.\ be able to find highly optimized IBP
reduction schemes using a small number of reduction runs.

In the future, we plan to publish our custom IBP reduction program
that provides step-by-step feedback on computational costs with a
standard interface (e.g.\ Ref.~\cite{towers2024gymnasium}), to
facilitate follow-up investigations by both physics and machine
learning communities. The present study focuses on examples that can
be quickly tested on a laptop in the matter of minutes, leaving many
possible future enhancements, e.g.\ more sophisticated neural network
architectures like convolutional neural networks, graph neural
networks and transformers. It would also be interesting to apply
reinforcement learning and metaheuristics to other problems in Feynman
integral evaluations besides IBP reduction.

\textbf{\textit{Acknowledgments---}} We thank Johann Usovitsch for
assistance in re-analyzing internal computation steps of
{\tt Kira} against our cost model to provide insightful comparisons
with AI results.  M.Z.’s work is supported in part by the U.K.\ Royal
Society through Grant URF\textbackslash R1\textbackslash 20109.  For
the purpose of open access, the authors have applied a Creative
Commons Attribution (CC BY) license to any Author Accepted Manuscript
version arising from this submission. We use the software {\tt
  Makie.jl} \cite{DanischKrumbiegel2021} for general visualization and
{\tt NN-SVG} \cite{lenail2019nn} for the neural network illustration.

\bibliography{main.bib}

\clearpage

\appendix
\onecolumngrid
\vspace{0.8cm}
\begin{center}
\textbf{Supplemental Material}
\end{center}

\begin{center}
\text{\textit{A. Algorithm implementation details}}
\end{center}

We use a private IBP reduction program written in the Julia language
\cite{bezanson2017julia}. The program allows using seed-operator pairs
in an arbitrary order and eliminating integrals in an arbitrary
order. After each step, the program can produce a representation of
the current state of IBP reduction, used as the observation in
reinforcement learning. The cost in our simplified model, roughly
corresponding to the number of arithmetic operations, can be returned
after each step (for use with reinforcement learning) or as a sum
after completing IBP reduction (for use with metaheuristics). The
program works with either rational parameters, e.g.\
Eq.~\eqref{eq:numericalKinematics}, used for $I_{3,3}$ reduction runs,
or finite-field parameters, used for $I_{50,50}$ runs.

We use the proximal policy optimization (PPO) algorithm
\cite{schulman2017proximal} for deep reinforcement learning,
implemented in the Julia package {\tt Crux.jl}
\cite{CruxPackage}. Besides the neural network for selecting actions,
i.e.\ the actor network described in the main text, a second neural
network, i.e.\ a critic network, is used to estimate the value of a
particular state and provide intermediate feedback for the quality of
the actions taken. The critic network has a similar MLP architecture
except that he final layer outputs a single number. The total number
of trainable parameters in the two neural networks is about 300
thousand. The activation function is the Gaussian error linear unit
(GELU) for the first hidden layer and the rectified linear unit (ReLU)
for the second and third hidden layers, in both networks. The two
neural networks are each trained for up to 20 epochs after every 1024
interactions with the environment, using the ADAM optimizer with a
learning rate of $2\times 10^{-4}$ and a mini-batch size of 256. The
information in Fig.~\ref{fig:observations} is embedded as a vector of
real numbers in the following manner. The upper triangle, lower
triangle and circle in each grid are each represented as a number, set
to zero if the feature is absent. A triangle or circle with step
number $n$ is embedded as $\cos(n/50)$, while an empty circle
(denoting an elimination candidate) is encoded as $-1$.  Some
PPO-specific hyperparameters are as follows.  The standard clip ratio
of 0.2 is used. An entropy bonus with a coefficient of 0.1 is added on
top of a normalized generalized advantage estimate (GAE) with a lambda
parameter of 0.95. The discount factor for rewards is 1, which is safe
since the environment always terminates in a finite number of
steps. Training for the actor network is terminated early if the
average KL divergence from the old action probabilities exceeds 0.012.

We use simulated annealing implemented by the Julia package {\tt
  Metaheuristics.jl} \cite{metaheuristics2022} which adapts an earlier
MATLAB code \cite{CorteSA}. An adaptive temperature is implemented by
the package: first, the inverse temperature rises linearly from 1 to
$10^4$ from the beginning to the end of the run to implement cooling;
second, the temperature at each step is dynamically rescaled with an
additional factor that is the absolute value of last function
evaluation, so that the user does not need to manually set a suitable
overall scale for the initial temperature. The tiny neural network
illustrated in Fig.~\ref{fig:nn} uses the GELU activation function for
the hidden layer. The final output does not have a bias parameter
since only the relative magnitudes of priority values matter, and the
network has a total of 15 parameters.

\begin{figure}[h]
    \centering
    \begin{subfigure}[b]{0.4\textwidth}
        \includegraphics[width=\textwidth]{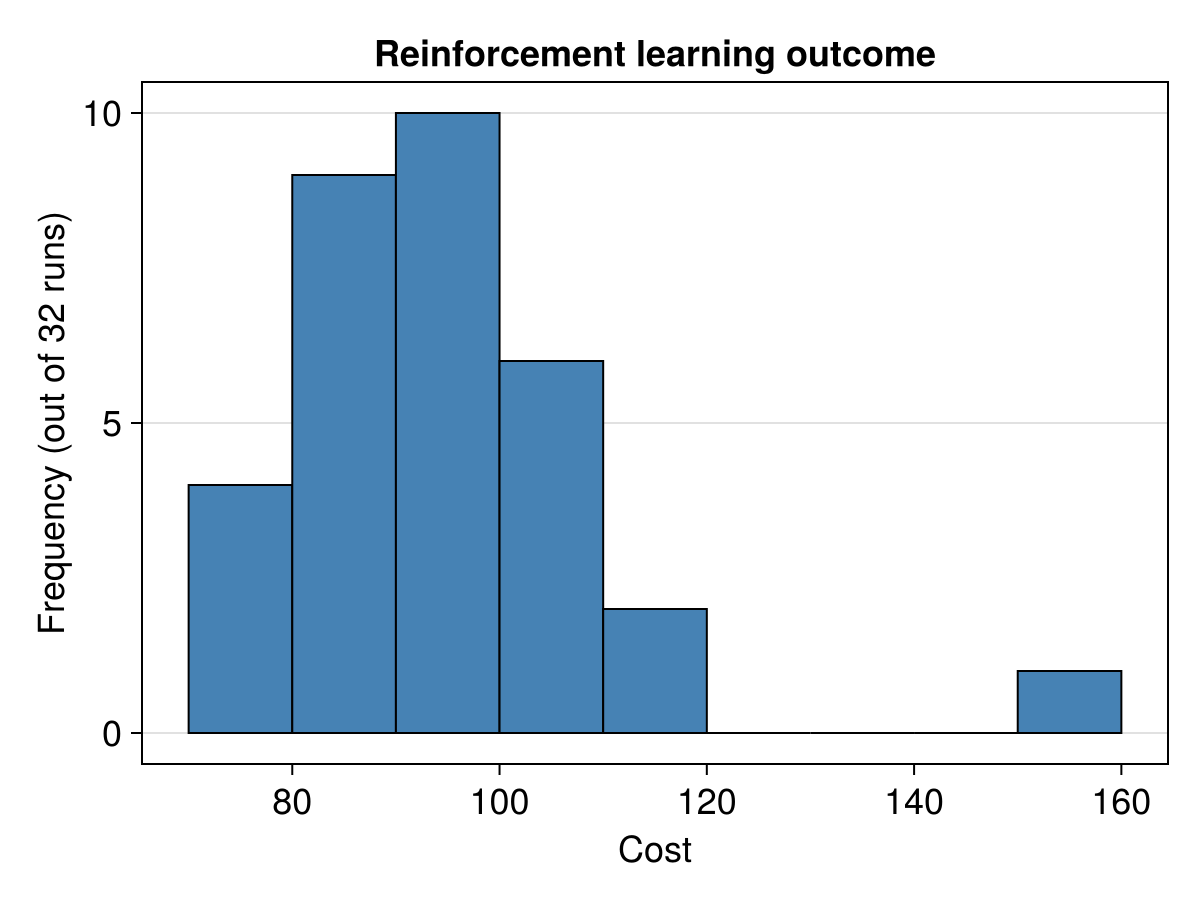}
        \caption{}
    \end{subfigure}
    \begin{subfigure}[b]{0.4\textwidth}
        \includegraphics[width=\textwidth]{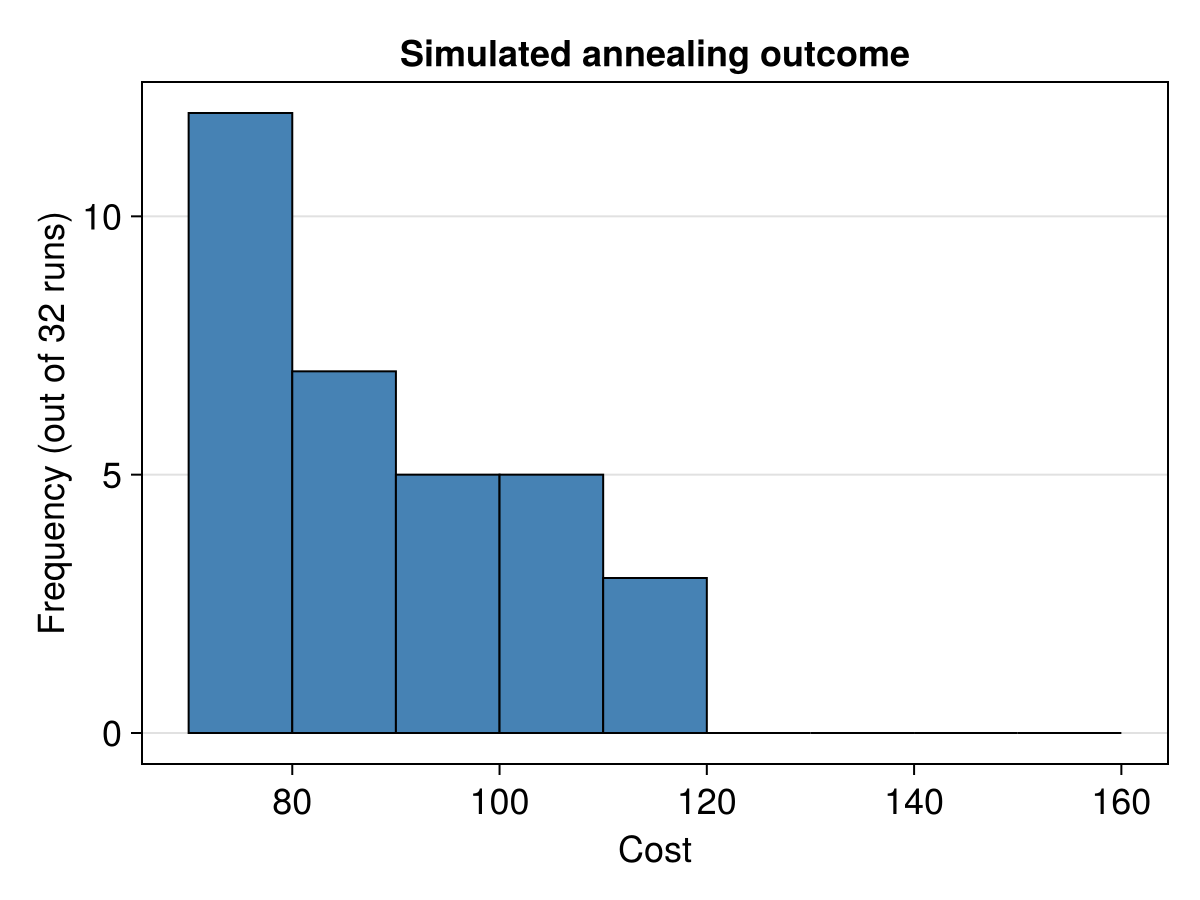}
        \caption{}
    \end{subfigure}
    \caption{Distribution of cost for reducing $I_{3,3}$ from 32 RL
      runs (left) and SA runs (right).}
    \label{fig:distribution}
\end{figure}

Each RL run takes about 5 minutes for $I_{3,3}$. This likely can be
made faster by training on GPU rather than CPU. Each SA run takes
about 5 seconds for $I_{3,3}$ and a few hours for $I_{50,50}$. The
full distribution of the cost achieved in RL and SA (without parameter
compression) for reducing $I_{3,3}$, in both cases with 32 runs each
with about 10000 complete IBP reductions, is shown in Fig.\
\ref{fig:distribution}.

\vspace{4mm}
\begin{center}
\text{\textit{B. Optimized reduction steps for $I_{3,3}$ found by
    reinforcement learning}}
\end{center}
Here we present the optimized IBP reduction steps for $I_{3,3}$,
corresponding to the graphical illustration Fig.~\ref{fig:i33best},
with 15 steps for generating equations and 15 steps for choosing
integrals to eliminate, with a total cost of 74 according to our
simplified cost model. The explicit form of the IBP equations from the two
IBP operators are:
\begin{itemize}
\item IBP operator 1:
  \begin{align}
    0 &= (D-2\nu_1-\nu_2) I_{\nu_1, \nu_2} + 2\nu_1 m^2 I_{\nu_1+1,
    \nu_2}  - \nu_2 I_{\nu_1-1, \nu_2+1} + \nu_2 (2m^2-p^2) I_{\nu_1,
      \nu_2+1} \, .
  \end{align}
\item IBP operator 2:
  \begin{align}
    0 &= (\nu_2-\nu_1) I_{\nu_1, \nu_2} + \nu_1 I_{\nu_1+1, \nu_2-1} -
        \nu_2 I_{\nu_1-1, \nu_2+1}  + \nu_1 p^2 I_{\nu_1+1, \nu_2} - \nu_2 p^2 I_{\nu_1,
      \nu_2+1} \, .
  \end{align}
  
The text below is machine-generated, adopting
the notation {\tt G[a,b]} to denote the bubble integral $I_{a,b}$. We
use the parameter values Eq.~\eqref{eq:numericalKinematics}, but we
have checked that the cost remains unchanged at other generic values
of the parameters.
\end{itemize}

{\small
\begin{verbatim}
Step 1: generate IBP equation from seed integral I[0,1] and IBP operator 1
IBP equation: (-2/7)*G[0,1] + (3)*G[0,2] + (-1)*G[-1,2] == 0
IBP equation reduced by previous rules: (-2/7)*G[0,1] + (3)*G[0,2] + (-1)*G[-1,2] == 0
reduction cost: 0

Step 2: new reduction rule from above equation: G[-1,2] -> (-2/7)*G[0,1] + (3)*G[0,2] 
normalization cost: 3

Step 3: generate IBP equation from seed integral I[3,2] and IBP operator 1
IBP equation: (-2)*G[2,3] + (-51/7)*G[3,2] + (6)*G[3,3] + (12)*G[4,2] == 0
IBP equation reduced by previous rules: (-2)*G[2,3] + (-51/7)*G[3,2] + (6)*G[3,3] + (12)*G[4,2] == 0
reduction cost: 0

Step 4: new reduction rule from above equation: G[4,2] -> (1/6)*G[2,3] + (17/28)*G[3,2] + (-1/2)*G[3,3] 
normalization cost: 4

Step 5: generate IBP equation from seed integral I[3,2] and IBP operator 2
IBP equation: (-2)*G[2,3] + (-1)*G[3,2] + (3)*G[4,1] + (-2)*G[3,3] + (3)*G[4,2] == 0
IBP equation reduced by previous rules: (-3/2)*G[2,3] + (23/28)*G[3,2] + (3)*G[4,1] + (-7/2)*G[3,3] == 0
reduction cost: 4

Step 6: new reduction rule from above equation: G[3,3] -> (-3/7)*G[2,3] + (23/98)*G[3,2] + (6/7)*G[4,1] 
normalization cost: 4

Step 7: generate IBP equation from seed integral I[3,1] and IBP operator 1
IBP equation: (-1)*G[2,2] + (-44/7)*G[3,1] + (3)*G[3,2] + (12)*G[4,1] == 0
IBP equation reduced by previous rules: (-1)*G[2,2] + (-44/7)*G[3,1] + (3)*G[3,2] + (12)*G[4,1] == 0
reduction cost: 0

Step 8: new reduction rule from above equation: G[4,1] -> (1/12)*G[2,2] + (11/21)*G[3,1] + (-1/4)*G[3,2] 
normalization cost: 4

Step 9: generate IBP equation from seed integral I[2,2] and IBP operator 2
IBP equation: (-2)*G[1,3] + (2)*G[3,1] + (-2)*G[2,3] + (2)*G[3,2] == 0
IBP equation reduced by previous rules: (-2)*G[1,3] + (2)*G[3,1] + (-2)*G[2,3] + (2)*G[3,2] == 0
reduction cost: 0

Step 10: new reduction rule from above equation: G[2,3] -> (-1)*G[1,3] + (1)*G[3,1] + (1)*G[3,2] 
normalization cost: 4

Step 11: generate IBP equation from seed integral I[2,2] and IBP operator 1
IBP equation: (-2)*G[1,3] + (-37/7)*G[2,2] + (6)*G[2,3] + (8)*G[3,2] == 0
IBP equation reduced by previous rules: (-8)*G[1,3] + (-37/7)*G[2,2] + (6)*G[3,1] + (14)*G[3,2] == 0
reduction cost: 4

Step 12: new reduction rule from above equation: G[3,2] -> (4/7)*G[1,3] + (37/98)*G[2,2] + (-3/7)*G[3,1] 
normalization cost: 4

Step 13: generate IBP equation from seed integral I[1,2] and IBP operator 1
IBP equation: (-2)*G[0,3] + (-23/7)*G[1,2] + (6)*G[1,3] + (4)*G[2,2] == 0
IBP equation reduced by previous rules: (-2)*G[0,3] + (-23/7)*G[1,2] + (6)*G[1,3] + (4)*G[2,2] == 0
reduction cost: 0

Step 14: new reduction rule from above equation: G[1,3] -> (1/3)*G[0,3] + (23/42)*G[1,2] + (-2/3)*G[2,2] 
normalization cost: 4

Step 15: generate IBP equation from seed integral I[2,1] and IBP operator 1
IBP equation: (-1)*G[1,2] + (-30/7)*G[2,1] + (3)*G[2,2] + (8)*G[3,1] == 0
IBP equation reduced by previous rules: (-1)*G[1,2] + (-30/7)*G[2,1] + (3)*G[2,2] + (8)*G[3,1] == 0
reduction cost: 0

Step 16: new reduction rule from above equation: G[3,1] -> (1/8)*G[1,2] + (15/28)*G[2,1] + (-3/8)*G[2,2] 
normalization cost: 4

Step 17: generate IBP equation from seed integral I[1,2] and IBP operator 2
IBP equation: (-2)*G[0,3] + (1)*G[1,2] + (1)*G[2,1] + (-2)*G[1,3] + (1)*G[2,2] == 0
IBP equation reduced by previous rules: (-8/3)*G[0,3] + (-2/21)*G[1,2] + (1)*G[2,1] + (7/3)*G[2,2] == 0
reduction cost: 4

Step 18: new reduction rule from above equation: G[2,2] -> (8/7)*G[0,3] + (2/49)*G[1,2] + (-3/7)*G[2,1] 
normalization cost: 4

Step 19: generate IBP equation from seed integral I[1,1] and IBP operator 2
IBP equation: (-1)*G[0,2] + (1)*G[2,0] + (-1)*G[1,2] + (1)*G[2,1] == 0
IBP equation reduced by previous rules: (-1)*G[0,2] + (1)*G[2,0] + (-1)*G[1,2] + (1)*G[2,1] == 0
reduction cost: 0

Step 20: new reduction rule from above equation: G[1,2] -> (-1)*G[0,2] + (1)*G[2,0] + (1)*G[2,1] 
normalization cost: 4

Step 21: generate IBP equation from seed integral I[0,2] and IBP operator 2
IBP equation: (2)*G[0,2] + (-2)*G[0,3] + (-2)*G[-1,3] == 0
IBP equation reduced by previous rules: (2)*G[0,2] + (-2)*G[0,3] + (-2)*G[-1,3] == 0
reduction cost: 0

Step 22: new reduction rule from above equation: G[-1,3] -> (1)*G[0,2] + (-1)*G[0,3] 
normalization cost: 3

Step 23: generate IBP equation from seed integral I[1,1] and IBP operator 1
IBP equation: (-1)*G[0,2] + (-16/7)*G[1,1] + (3)*G[1,2] + (4)*G[2,1] == 0
IBP equation reduced by previous rules: (-4)*G[0,2] + (3)*G[2,0] + (-16/7)*G[1,1] + (7)*G[2,1] == 0
reduction cost: 4

Step 24: new reduction rule from above equation: G[2,1] -> (4/7)*G[0,2] + (-3/7)*G[2,0] + (16/49)*G[1,1] 
normalization cost: 4

Step 25: generate IBP equation from seed integral I[0,2] and IBP operator 1
IBP equation: (-9/7)*G[0,2] + (6)*G[0,3] + (-2)*G[-1,3] == 0
IBP equation reduced by previous rules: (-23/7)*G[0,2] + (8)*G[0,3] == 0
reduction cost: 3

Step 26: new reduction rule from above equation: G[0,3] -> (23/56)*G[0,2] 
normalization cost: 2

Step 27: generate IBP equation from seed integral I[1,0] and IBP operator 1
IBP equation: (-9/7)*G[1,0] + (4)*G[2,0] == 0
IBP equation reduced by previous rules: (-9/7)*G[1,0] + (4)*G[2,0] == 0
reduction cost: 0

Step 28: new reduction rule from above equation: G[2,0] -> (9/28)*G[1,0] 
normalization cost: 2

Step 29: generate IBP equation from seed integral I[0,1] and IBP operator 2
IBP equation: (1)*G[0,1] + (-1)*G[0,2] + (-1)*G[-1,2] == 0
IBP equation reduced by previous rules: (9/7)*G[0,1] + (-4)*G[0,2] == 0
reduction cost: 3

Step 30: new reduction rule from above equation: G[0,2] -> (9/28)*G[0,1] 
normalization cost: 2
\end{verbatim}
}

\end{document}